\documentclass[11pt]{article}
\pdfoutput=1
\newcommand{\Comment}[1]{{}}
\usepackage{amsfonts,amsthm,amsmath,amssymb,slashed}
\usepackage[textwidth = 430 pt, textheight = 630 pt]{geometry}
\usepackage{color}

\usepackage{bm}
\usepackage{cite}
\usepackage{mathrsfs}
\usepackage{graphicx}
\usepackage{verbatim}
\usepackage[dvipsnames]{xcolor}
\usepackage{enumerate}
\usepackage{authblk}
\usepackage[hang,flushmargin]{footmisc}
\usepackage{lipsum}
\usepackage[font=footnotesize]{caption}
\usepackage{soul}
\usepackage{empheq}
\usepackage{cancel}

\Comment{\usepackage{color}
\definecolor{MyDarkBlue}{rgb}{0.15,0.15,0.45}
\usepackage[linktocpage=true]{hyperref}
\hypersetup{
colorlinks=true,
citecolor=MyDarkBlue,
linkcolor=MyDarkBlue,
urlcolor=MyDarkBlue,
pdfauthor={Horatiu Nastase, Francisco Rojas and Carlos Rubio},
pdftitle={Celestial IR divergences in general most-subleading-color gluon and gravityamplitudes},
pdfsubject={hep-th}
}
\usepackage[utf8]{inputenc}
\usepackage[numbers,sort&compress]{natbib}
\usepackage{hypernat}}
\usepackage{graphicx}
\usepackage{cite}
\usepackage[colorlinks=true]{hyperref}
\hypersetup{
  allcolors = blue,
}

\newcommand\ignore[1]{}
\def\one{{\,\hbox{1\kern-.8mm l}}}

\def\ket#1{\left| #1\right\rangle}

\def\Tr{{\rm Tr\, }}

\def\a{\alpha}\def\b{\beta}

\def\d{\partial}

\def\Tr{\mathop{\rm Tr}\nolimits}

\newcommand{\Cset}{{\,\,{{{^{_{\pmb{\mid}}}}\kern-.45em{\mathrm C}}}}}

\newcommand{\cA}{\mathcal A}
\newcommand{\cB}{\mathcal B}

\newcommand{\cJ}{\mathcal J}
\newcommand{\cM}{\mathcal M}
\newcommand{\cN}{\mathcal N}\newcommand{\cO}{\mathcal O}
\newcommand{\cP}{\mathcal P}
\newcommand{\cR}{\mathcal R}\newcommand{\cS}{\mathcal S}

\newcommand{\be}{\begin{equation}}
\newcommand{\bea}{\begin{eqnarray}}

\newcommand{\ee}{\end{equation}}
\newcommand{\eea}{\end{eqnarray}}

\newcommand{\nn}{\nonumber}

\def\bT{\mathbf{T}}

\def\bS{\mathbf{S}}

\def\bM{\mathbf{M}}
\def\bGam{\mathbf{\Gamma}}

\def\eps{ \epsilon }

\def\beann{\begin{eqnarray*}}
\def\eeann{\end{eqnarray*}}
\def\beq{\begin{equation}}
\def\eeq{\end{equation}}
\def\ba{\begin{array}}
\def\ea{\end{array}}
\def\ben{\begin{enumerate}}

\def\bA{\boldsymbol{A}}

\def\bZ{{\bf Z}}
\def\bz{{\bar z}}

\def\ones{\left( \begin{array}{c} 1 \\ 1 \\ 1 \\ \end{array} \right) }
\def\Tete{{(2\ell,2 \ell)}}

\def\Teptep{{(2\ell+1,2 \ell+1)}}

\def\lam{\lambda}

\def\eps{\epsilon} 
\def\ep{\epsilon}

\def\mommu{ { s_{ij} \over \mu^2 } }
\def\as{a(\mu^2)}
\def\bas{\bar{a}}

\def\Ell{{(L)}}

\def\Ellk{{(L,k)}}
\def\Ellodd{{(2\ell+1,2\ell+1)}}
\def\Elleven{{(2\ell,2\ell)}}

\def\Zero{{(0)}}
\def\One{{(1)}}

\def\Oneone{{(1,1)}}

\def\Three{{(3)}}

\def\Threethree{{(3,3)}}

\newcommand{\eal}[1]{\begin{equation} \begin{aligned} #1 \end{aligned}\end{equation}}
\newcommand{\lr}[1]{\langle #1 \rangle}

 \newcommand{\badat}{\begin{alignedat}}
 \newcommand{\eadat}{\end{alignedat}}

\begin{document}

\renewcommand{\thefootnote}{\fnsymbol{footnote}}

\makeatletter
\@addtoreset{equation}{section}
\makeatother
\renewcommand{\theequation}{\thesection.\arabic{equation}}

\rightline{}
\rightline{}




\begin{center}
{\LARGE \bf{\sc Celestial IR divergences in general\\[-3pt]most-subleading-color gluon\\[5pt] and gravity amplitudes }}
\end{center}
 \vspace{1truecm}
\thispagestyle{empty} \centerline{
{\large \bf {\sc Horatiu Nastase${}^{a}$}}\footnote{E-mail address: \Comment{\href{mailto:horatiu.nastase@unesp.br}}{\tt 
horatiu.nastase@unesp.br}},
{\large \bf {\sc Francisco Rojas${}^{b}$}}\footnote{E-mail address: \Comment{\href{mailto:francisco.rojasf@uai.cl}}{\tt francisco.rojasf@uai.cl}}
{\bf{\sc and}}
{\large \bf {\sc Carlos Rubio${}^{b}$}}\footnote{E-mail address: \Comment{\href{mailto:carlos.rubio@edu.uai.cl}}{\tt carlos.rubio@edu.uai.cl}}
}

\vspace{.5cm}

\centerline{{\it ${}^a$Instituto de F\'{i}sica Te\'{o}rica, UNESP-Universidade Estadual Paulista}}
\centerline{{\it R. Dr. Bento T. Ferraz 271, Bl. II, Sao Paulo 01140-070, SP, Brazil}}

\vspace{.3cm}

\centerline{{\it${}^b$Facultad de Ingenier\'ia y Ciencias, Universidad Adolfo Ib\'a\~nez}}
\centerline{{\it Diagonal Las Torres 2640, Pe\~nalol\'en, Santiago, Chile}}

\vspace{1truecm}

\thispagestyle{empty}

\centerline{\sc Abstract}

\vspace{.4truecm}

\begin{center}
\begin{minipage}[c]{450pt}
{\noindent 
Gluon amplitudes at most-subleading order in the $1/N$ expansion share a remarkable simplicity with graviton amplitudes: collinear 
divergences are completely absent in both and, as a consequence, their full IR behavior arises from soft gluon/graviton exchange 
among the external states. 
In this paper we study the effect of all-loop IR divergences of celestial most-subleading color gluon amplitudes 
and their similarities with the celestial gravity case. In particular, a simple celestial exponentiation formula for the 
dipole part can be written. We also analize how this exponentiation is modified by non-dipole contributions.
Finally we also show that, in the Regge limit, the soft factor satisfies the Knizhnik-Zamolodchikov equation hinting 
at the possibility that, in this limit, an effective Wess-Zumino-Witten model would describe the dynamics of the infrared sector. 
}
\end{minipage}
\end{center}

\vspace{.5cm}

\setcounter{page}{0}
\setcounter{tocdepth}{2}

\newpage

\renewcommand{\thefootnote}{\arabic{footnote}}
\setcounter{footnote}{0}

\linespread{1.1}
\parskip 4pt




\tableofcontents

\section{Introduction}
Celestial amplitudes offer an explicit realization of flat space holography. Momentum space amplitudes for massless 
external states are mapped to correlation functions on the two-dimensional celestial sphere at the null boundary of Minkowski space. 
After a Mellin transform over the energies of all the external particles participating in the scattering event, the Lorentz 
group SL(2,$\mathbb{C}$) identifies the plane-wave asymptotic states with conformal primary fields defined on the 
celestial sphere at null infinity \cite{Cheung:2016iub,Pasterski:2017ylz,Pasterski:2017kqt}. 
Celestial amplitudes thus exhibit the usual tranformation properties of correlation 
functions of a two-dimensional CFT, however, the precise underlying conformal field theory (or theories) 
describing these correlators is still largely unknown.

Momentum space amplitudes in gravity and gauge theories are plagued with infrared divergences. These infinities 
appear due to the interaction of external particles with an infinite amount of photons, gluons or gravitons carrying 
arbitrarily low energies. Remarkably, the entire effect of infrared divergences can be fully traced yielding a 
factorized form for the $n$-point amplitude, namely 
\be
\cA_n = \bZ_n \, \cA_{\rm hard}\,,
\ee
where the entirety of the infrared divergent part is contained in the \emph{universal} soft factor $\bZ_n$, 
whereas $\cA_{\rm hard}$ is a process-dependent (and infrared-safe) $n$-particle amplitude. 

It has recently been shown that this factorization persists for celestial amplitudes 
\cite{Gonzalez:2020tpi,Arkani-Hamed:2020gyp,Gonzalez:2021dxw}, with $\bZ$ being a function that depends 
solely on the celestial coordinates ${z_i}$. For example, the Mandelstam invariants $s_{ij}$ get mapped to 
``squared distances'' $|z_{ij}|^2$ on the celestial sphere. Thus, being this factor a universal quantity depending only on celestial 
data, it appears to be a perfect candidate to describe general features of the precise but yet undiscovered celestial CFT (CCFT). 
In particular, one would like, at least approximately, to answer the question: What is the holographic two-dimensional CFT 
living on the celestial sphere at null infinity governing the infrared degrees of freedom of the four-dimensional gauge theory in the bulk? 

Up until now, many steps have been taken in this direction. In \cite{Kalyanapuram:2020epb} a theory of free scalars 
was put forward in order to describe the infrared sector of abelian gauge theory and graviton amplitudes. 
In \cite{Magnea:2021fvy,Gonzalez:2021dxw} an analogous theory of free colored scalars was proposed in 
order to reproduce the celestial correlation function given by the celestial soft factor $\bZ$ in nonabelian 
gauge theory. While succesfully describing the infrared sectors in terms of a 2-dimensional description of free scalar 
fields, we know that this is far from being the end of the story, since nonabelian gauge theory is much richer and subtle. 
For instance in \cite{Magnea:2021fvy}, the 2-dimensional theory takes into account the theory at finite $N$, but 
considering only the dipole contributions to the soft anomalous dimension matrix. In \cite{Gonzalez:2021dxw}, 
all (known) contributions to the soft anomalous dimension beyond the dipole approximation were considered, but 
only in the large $N$ limit. In both these cases, and in the abelian construction in \cite{Kalyanapuram:2020epb}, 
one may suspect that the absence of the effects from collinear divergences may be responsible for the apparent 
simplicity of the holographic description obtained \cite{Magnea}. Therefore, we believe that a richer 2-dimensional theory, beyond 
free colored-scalars, must be at play governing the dynamics of the infrared sector of nonabelian gauge 
theory.\footnote{Very recently, a non-local theory has been proposed describing the infrared sector in abelian 
gauge theory and gravity in higher spacetime dimensions \cite{Kapec:2021eug}.}  

In this article we take a step in this direction by focusing on the sub-sector of gluon amplitudes called 
most-subleading-color amplitudes 
\cite{Naculich:2008ys,Naculich:2009cv,Nastase:2010xa,Naculich:2011pd,Naculich:2011my,Naculich:2013xa}. 
These amplitudes are a specific (infinite) subset of loop contributions
to the full amplitude that are most-subleading in the $1/N$ expansion at each order in the strong coupling 
constant. One of its remarkable features is that, similar to graviton amplitudes, these amplitudes are free of 
collinear divergences. This makes its infrared structure remarkably simple, while still being at finite $N$. 
In fact, within the dipole approximation, the infrared divergences are one-loop exact, and can be explicitly 
computed for a general SU($N$) gauge theory \cite{Naculich:2013xa}.

The similarity between graviton and most-subleading-color gluon amplitudes and their infrared structure is not a coincidence. There are 
a number of relations between them: exact relations for the 4-point functions at $L=0,1,2$ loops, valid 
not only for the maximal supersymmetric cases (${\cal N}=8$ 
vs. ${\cal N}=4$)\cite{Naculich:2008ys}, but also for arbitrary ${\cal N}$ \cite{Naculich:2011my}, exact relations at one-loop 5-points 
extended from maximal supersymmetry \cite{Nastase:2010xa} to general ${\cal N}$ \cite{Naculich:2011my}, 
and relations between the first two IR-divergent terms in the maximal supersymmetry case  \cite{Naculich:2013xa}.

It is important to note that these relations among graviton and most-subleading-color gluon amplitudes hold 
for the full amplitudes, that is, soft and hard parts included. However, up until now, these connections are 
only available in a loop-by-loop analysis. This means that, contrary to the soft factor whose basic structure 
is known to all-loop orders, we do not yet have the luxury of making a statement between the celestial correlators 
for the full amplitudes of these theories. With this is mind, in this article we perform a loop order analysis for the 
celestial loop amplitudes similar to that of planar $\cN=4$ SYM performed in \cite{Gonzalez:2020tpi}. 

We will also see that their infrared properties, being similar to those of gravity, makes writing down the celestial factorization of 
infrared divergences almost trivial in this case. Finally, we will focus on the celestial soft factor and show that, 
in the forward limit (Regge limit), it satisfies the Knizhnik-Zamolodchikov equation, signalling a WZW model as 
an effective theory of the infrared degrees of freedom in the Regge limit. 

This paper is organized as follows. In section 2 we review celestial loop amplitudes and their factorization properties. 
In section 3 we consider celestial graviton amplitudes and their infrared divergent structure, with a view towards their 
relation with most-subleading gluon amplitudes. 
In section 4 we focus on the case of most-subleading gluon amplitudes in SU($N$) gauge theory, studying their 
infrared factorization properties and their celestial avatar. We will restrict ourselves to dipole contributions only in 
the soft anomalous dimension matrix. Then, in section 5, we will see how the Knizhnik-Zamolodchikov equation 
emerges, in the Regge limit, as a consistency condition for the soft factor in the most-subleading sector. Finally, 
in section 6 we analyze corrections to our results beyond the dipole approximation.

\section{Celestial amplitudes, ${\cal N}=4 $ SYM and IR divergences}
In this section we review the general results obtained in \cite{Gonzalez:2020tpi} for the exact 4-point celestial 
amplitude with massless external states, and the structure of infrared divergences at arbitrary loop order in the 
planar limit of $\mathcal{N}=4$ SYM theory.
\subsection{Celestial amplitudes}
\label{sec:Setup}

We consider scattering amplitudes in four dimensions\footnote{Here (in this section) we work in $(+---)$ signature.} with external 
massless particles. Each external state is labelled by a momentum $p_i^\mu$, a helicity $\ell_i$, and a sign 
$\eta_i=\pm 1$ distinguishing incoming ($-1$) from ($+1$) outgoing states. We parameterize $p_i^\mu$ in 
terms of points $(z_i,\bar{z}_i)$ on the two-dimensional celestial sphere through the map
\be
\label{p}
 p_i^\mu=\frac{1}{2} \eta_i \omega_i \left(1+|z_i|^2, z_i+\bar{z}_i,-i(z_i-\bar{z}_i),1-|z_i|^2 \right)\,,
\ee
with $\omega_i\geq0$.

\subsubsection{Basis for 4-point amplitudes}
\label{ssec:HelicityBasis}

With the external particles written in terms of their helicities $\ell_i$, momenta $p_i$ as in \eqref{p}, and
stripping off the overall momentum-conserving delta function, the momentum-space amplitude is denoted as
\begin{equation}
\label{eq:4ptscattering}
 \cA\left(\{\omega_i,\ell_i,z_i,\bar{z}_i\}\right)=A\left(1^{\ell_1},2^{\ell_2},3^{\ell_3},4^{\ell_4}\right)   \delta(p_1+p_2-p_3-p_4) \,.
\end{equation}

It is also convenient to further split the stripped amplitude as 
\be
\label{gen4pt}
A\left(1^{\ell_1},2^{\ell_2},3^{\ell_3},4^{\ell_4}\right)=  \cB(s,t) \; \cR_{(\ell_1,\ell_2,\ell_3,\ell_4)}\,,
\ee
where $\cB(s,t)$ is a function of two of the Mandelstam variables $s=(p_1+p_2)^2$, $t=(p_1-p_4)^2$, and 
$u=(p_1-p_3)^2$ (which is not necessarily an analytic function), and the full dependence on the helicities of 
the external particles is contained in the rational function $\cR_{(\ell_1,\ell_2,\ell_3,\ell_4)}$. The latter can be 
written in terms of the usual spinor helicity products $\langle i j\rangle$ 
and $[i j]$ which, rewritten in terms of  the complex variables $(z,\bar z)$ on the 
celestial sphere and the energy 
$\omega$, become
\be
\label{prod}
\langle i j \rangle= \epsilon_{\alpha \beta} \lambda^{\alpha}_{i} \lambda^{\beta}_{j}= \sqrt{\omega_i \omega_j} z_{ij}\,,\quad 
[ i j ]= - \epsilon_{\dot \alpha \dot \beta} \tilde{\lambda}^{\dot \alpha}_{i} \tilde{\lambda}^{\dot \beta}_{j}
= - \sqrt{\omega_i \omega_j} \bar{z}_{ij}\,,
\ee
with $z_{ij}=z_i-z_j$ and $\bar{z}_{ij}=\bar{z}_i-\bar{z}_j$. The rational function $\cR_{(\ell_1,\ell_2,\ell_3,\ell_4)}$ 
can also have an explicit dependence on the SL(2,$\mathbb{C}$) invariant cross-ratio
\be
\label{eq:crossratio}
r=\frac{z_{12} z_{34}}{z_{23} z_{41}}\,,
\ee 
which, using total momentum conservation, is directly related to the scattering angle $\theta_{cm}$ in the center-of-mass frame through 
\eal{\label{eq:rst}
r=-\frac{s}{t} = \csc^2\left(\frac{\theta_{cm}}{2}\right)\,.
}
from where we see that $r$ is real and $r>1$. 

The Regge (or forward) limit corresponds to taking large $r$ and we will 
particularly focus in this regime in Section \ref{sec:KZ}. Also, although from \eqref{eq:crossratio} it may seem that $r$ 
is complex, total momentum conservation forces $r$ to be real (see, for instance, equation \eqref{delta}), which is 
simply a consequence of the fact that the four-point amplitude always occurs on a plane. 

As all the information about the helicities $\ell_i$ of the external states is encoded in the function $\cR_{(\ell_1,\ell_2,\ell_3,\ell_4)}$, 
the function must be an eigenfunction of the $\hat \ell_i$,
\be \label{hEVeq}
\hat{ \ell}_i \cR_{(\ell_1,\ell_2,\ell_3,\ell_4)} = \ell_i \cR_{(\ell_1,\ell_2,\ell_3,\ell_4)}\,,
\ee 
where $\hat{\ell}_i$ is the helicity operator for the $i$-th particle,
\eal{
\hat{ \ell}_i = \frac{1}{2}\left(-\lambda^{\alpha}_i \frac{\partial}{\partial \lambda^{\alpha}_i } + 
\tilde{\lambda}^{\dot{\alpha}}_i \frac{\partial}{\partial \tilde{\lambda}^{\dot{\alpha}}_i } \right)\,.
}

We may express a solution to the eigenfunction constraint \eqref{hEVeq}  
in terms of powers of Lorentz invariant functions $R_i$, defined by
the relation $\hat{ \ell}_i R_j=\delta_{ij} R_j $, as
\be
\cR_{(\ell_1,\ell_2,\ell_3,\ell_4)}= r^{\alpha_1} (r-1)^{\alpha_2}  \prod^{4}_{i=1}R_i^{\ell_i}\,,
\ee
with $\alpha_1$ and $\alpha_2$ real numbers. The basis of functions $R_i$ we use is \cite{Gonzalez:2020tpi}
\begin{align}
\begin{split}
R_1=\left(\tfrac{[12][13] \langle23\rangle}{\langle 12\rangle \langle13 \rangle  [32]}\right)^{\frac{1}{2}}\,,&\quad R_2
=\left(\tfrac{[12][23] \langle13\rangle}{\langle 12\rangle \langle2 3 \rangle  [31]}\right)^{\frac{1}{2}}\,,\\
R_3= \left(\tfrac{[13][23] \langle12\rangle}{\langle 13\rangle \langle 23 \rangle  [21]}\right)^{\frac{1}{2}}\,,&\quad R_4
= \tfrac{[24]}{\langle 24 \rangle} \left(\tfrac{[31] \langle12\rangle \langle23\rangle}{ \langle 13\rangle  [12] [23]}\right)^{\frac{1}{2}}\,,
\end{split}
\end{align}
which only depends on the \emph{distances} $(z_{ij}, \bar{z}_{ij})$ on the celestial sphere.\footnote{For our purposes, 
this is a convenient basis because each $R_i$ is independent on the energies of the external states $\omega_i$. See 
\cite{Badger:2016uuq} for a different basis in which there is an explicit energy dependence.}

Then, the four-point stripped amplitude \eqref{gen4pt} can be expressed solely in terms of the (a priori) arbitrary function 
of Mandelstam variables $\cB(s,t)$ as 
\begin{multline}
\label{eq:4point}
A\left(1^{\ell_1},2^{\ell_2},3^{\ell_3},4^{\ell_4}\right)=r^{\alpha_1} (r-1)^{\alpha_2} \cB(s,t) \left(\tfrac{ z_{12} }
{\bar{z}_{12} }\right)^{-\tfrac{1}{2}(\ell_1+\ell_2-\ell_3-\ell_4)} \left(\tfrac{ z_{13} }{\bar{z}_{13} }\right)^{-\tfrac{1}{2}
(\ell_1-\ell_2+\ell_3+\ell_4)}\\ \times \left(\tfrac{ z_{23} }{\bar{z}_{23} }\right)^{\tfrac{1}{2}(\ell_1-\ell_2-\ell_3+\ell_4)}
\left(\tfrac{ z_{24} }{\bar{z}_{24} }\right)^{-\ell_4} \,.
\end{multline}

Notice that no mention whatsoever has been made to perturbation theory, thus, the statements made so far apply 
to the {\em exact} (all-loop) four-point $\cS$-matrix element.

\subsubsection{General 4-point celestial amplitude}\label{ssec:CelestialAmplitudes}

Celestial amplitudes are defined through a Mellin transform over the energies $\omega_i$ of each of the $n$ external 
states in a momentum-space amplitude $ \cA\left(\{\omega_i,\ell_i,z_i,\bar{z}_i\}\right)$, maintaining the variables 
$(z_i,\bar z_i)$ as coordinates on the celestial sphere, and reinterpreting the helicities $l_i$ as spins $J_i$ of operators,
\be
\label{eq:4}
\tilde{\cA}\left(\{\Delta_i,J_i,z_i,\bar{z}_i\}\right)= \prod^n_{k=1} \left( \int^\infty_0  d\omega_k 
\omega^{\Delta_k-1}_{k}\right) \cA\left(\{\omega_i,\ell_i,z_i,\bar{z}_i\}\right)\,.
\ee

After performing the transform, $\Delta_j$ are the conformal dimensions and $J_j \equiv \ell_j$ the spins of conformal 
primary operators inserted at the points $(z_j,\bz_j)\in S^2$. Indeed, under $SL(2,\mathbb{C})$ Lorentz transformations 
on the conformal sphere in $(z,\bar z)$ coordinates, the celestial amplitude \eqref{eq:4} transforms 
as~\cite{Pasterski:2017ylz,Pasterski:2017kqt,Stieberger:2018edy}
\begin{equation}
  \widetilde{\cA}\Big(\Big\{\Delta_j, J_i;\frac{a z_i+b}{cz_i+d},\frac{ \bar{a}  \bz_i+ \bar{b}}{ \bar{c} \bz_i
  + \bar{d}}\Big\}\Big) =\prod_{j=1}^n \Big( (c z_j+d)^{\Delta_j+J_j} (\bar{c} \bz_j+\bar{d})^{\Delta_j-J_j}\Big) 
  \widetilde{\cA}(\{\Delta_i,J_i;z_i,\bz_i\})\,.
\end{equation}
\emph{i.e.}, it has the correct transformation properties of a 2D CFT correlation function defined on the celestial sphere. 
Specializing to the 4-point case, inserting \eqref{eq:4ptscattering} with \eqref{gen4pt} into \eqref{eq:4}, and using the delta 
function expressed in terms of the celestial sphere, $(\omega_i,z_i,\bar z_i)$
\begin{multline}
\label{delta}
 \delta^{(4)}(p_1+p_2-p_3-p_4)=\frac{4}{\omega_4 |z_{14}|^2 |z_{23}|^2} \delta(r-\bar{r}) 
\\
  \times \delta\left(\omega_1 - \frac{z_{24}\bz_{34}}{z_{12}\bz_{13}}\omega_4\right)\delta\left(\omega_2 
  + \frac{z_{14}\bz_{34}}{z_{12}\bz_{23}}\omega_4\right)
 \delta\left(\omega_3 + \frac{z_{24}\bz_{14}}{z_{23}\bz_{13}}\omega_4\right)\,,
\end{multline}
we obtain the four-point correlator (up to a numerical factor)
\be
\label{eq:Celestial4pt}
\tilde{\cA}\left(\{\Delta_i,J_i,z_i,\bar{z}_i\}\right)= f(r,\bar{r}) \prod_{i<j}^4 z_{ij}^{\frac{h}{3}-h_i-h_j}
\bar{z}_{ij}^{\frac{\bar{h}}{3}-\bar{h}_i-\bar{h}_j}\,.
\ee

Here $r$ is the conformally invariant cross-ratio (\ref{eq:crossratio}) which, \emph{a priori}, is a complex number. 
However, because of the delta function $\delta(r-\bar r)$, the cross-ratio becomes real and equal to $r=-s/t$ due 
to total momentum conservation. The conformally invariant function $f(r,\bar{r})$ is given by
\be
\label{eq:f}
f(r,\bar{r}) =2  \delta(r-\bar{r}) \Theta(r-1) r^{\alpha_1+\frac{\Delta}{6}} (r-1)^{\alpha_2+\frac{\Delta}{6}} 
\int^{\infty}_0 dw\, w^{\tfrac{\Delta-6}{2} } \cB(r w, -w)\,,
\ee
with $\Delta \equiv \sum_{i=1}^4 \Delta_i$ the total conformal dimension. 

The conformal weights $(h_i,\bar h_i)$ in~\eqref{eq:Celestial4pt} are given in terms of scaling dimensions and conformal spins by 
\be
\label{eq:hs}
h_k=\frac{1}{2}(\Delta_k + J_k)\,, \quad \bar{h}_{k}=\frac{1}{2}(\Delta_k - J_k)\,,
\ee
where the scaling dimensions $\Delta_k$ are restricted to lie on the principal continuous series of the 
$SL(2,\mathbb{C})$ Lorentz group\cite{Pasterski:2017kqt}, \emph{i.e.},
\begin{equation}
 \Delta_k=1+i\lambda_k \;\; \text{with} \; \lambda_k\in \mathbb{R}.
\end{equation}

\label{ssec:TreeExample}

As an example, consider the celestial tree-level four-gluon amplitude in pure Yang-Mills\footnote{At tree level, the pure 
Yang-Mills gluon amplitude is the same amplitude as in any other ${\cal N}$ supersymmetric Yang-Mills theory, such as ${\cal N}=4$ SYM 
which is one of the cases of interest in this article.} first discussed in~\cite{Pasterski:2017ylz,Stieberger:2018edy}. 
The stripped MHV four-gluon amplitude is
\eal{\label{Atree}
A_{\rm tree}(1^-,2^-,3^+,4^+)&= g^2 \frac{\langle 12 \rangle^3}{\lr{23}\lr{34}\lr{41}}= g^2 \, r \frac{z_{12}\bar{z}_{34}}{\bar{z}_{12} z_{34}}\,,
}
where in the last step, the constraints from momentum conservation~\eqref{delta} were used. For the helicities 
$\ell_3=\ell_4=-\ell_1=-\ell_2=1$, from~\eqref{eq:4point} we read off that $\alpha_1=1$, $\alpha_2=0$, and the 
arbitrary function $\cB(s,t)= g^2$ is simply the coupling constant squared.\footnote{Hereafter we omit an overall 
factor of $i(2\pi)^4$.} 

After applying the Mellin transforms in \eqref{eq:4}, one obtains the four-point 
correlation function~\eqref{eq:Celestial4pt} with conformal weights are $(h_k,\bar{h}_k)
=(\frac{i}{2} \lambda_k,1+\frac{i}{2}\lambda_k)$ for the negative helicity gluons and $(h_k,\bar{h}_k)=(1+\frac{i}{2} 
\lambda_k,\frac{i}{2}\lambda_k)$ for the positive helicity ones. The conformally invariant factor $f(r,\bar r)$ is 
\eal{\label{eq:f_4gluontree}
f_{\rm tree}(r,\bar{r}) =2 g^2  \delta(r-\bar{r}) \Theta(r-1)r^{1+\frac{\Delta}{6}}(r-1)^{\frac{\Delta}{6}} \mathcal{I}(\lambda)\,,
}
where
\begin{equation}\label{eq:deltalambda}
 \mathcal{I}(\lambda)\equiv \int_0^\infty \frac{dw}{w}\, w^{i\frac{\lambda}{2}} =4\pi \delta(\lambda)\,,
\end{equation}
with $\lambda=\sum_{k=1}^4 \lambda_k$ being the imaginary part of $\Delta$. 
Enforcing the delta function, \eqref{eq:deltalambda} yields $\Delta=4$ giving
\begin{equation}
 f_{\rm tree}(r,\bar{r}) =8\pi g^2  \delta(r-\bar{r}) \Theta(r-1)r^{\frac{5}{3}}(r-1)^{\frac{2}{3}} \delta(\lambda)\,.
\end{equation}

\subsection{Loop amplitudes in ${\cal N}=4$ SYM and IR divergences}

Here we focus on gluon amplitudes in the planar limit of ${\cal N}=4$ SYM. Because this is a conformal field theory, 
scattering amplitudes would be ill-defined in $D=4$ dimensions. However, the need for introducing a regulator to 
deal with infrared divergences breaks conformal invariance and renders the amplitudes well defined. 
More specifically, we consider loop momenta in 
$D=4-2\epsilon$ dimensions while the momentum and polarization of the external particles remain in 
$D=4$ dimensions. Infrared divergences thus arise as poles in $\epsilon$ with $1/\epsilon^{2L}$ at $L$ 
loops.\footnote{We could of course consider other schemes to regulate infrared divergences, such as 
photon/gluon mass or momentum cut-off, but 
we will use dimensional regularization throughout this paper.}

The MHV all-loop 4-gluon amplitudes can be written as the tree amplitude, times a scalar factor expanded in a loop
expansion via the rescaled 't Hooft coupling $a$,
\bea
A&=& A_{\rm tree} M_\epsilon\cr
M_\epsilon&=&1+\sum_{L=1}^\infty a^L M_\epsilon^{(L)}\cr
a&=&\frac{g^2N}{8\pi^2}(4\pi e^{-\gamma_E})^\epsilon.
\eea

The BDS ansatz for the full rescaled 4-point amplitude $M_\epsilon$ can be written as \cite{Bern:2005iz,Alday:2007hr}
\bea
M_\epsilon&=&\left(A_\epsilon(s)\right)^2\left(A_\epsilon(t)\right)^2\exp\left\{\frac{f(\lambda)}{8}\left(\log^2(s/t)+\frac{\pi^2}{3}\right)
+C(\lambda)\right\}\cr
A_\epsilon(s)&=&\exp\left[-\frac{1}{8\epsilon^2}f^{(-2)}\left(\lambda\left(\frac{\mu^2}{-s}\right)^\epsilon\right)
-\frac{1}{4\epsilon}g^{(-1)}\left(\lambda\left(\frac{\mu^2}{-s}\right)^\epsilon\right)\right]\cr
f(\lambda)&\equiv& \left(\lambda\frac{d}{d\lambda}\right)^2f^{(-2)}(\lambda)\cr
g(\lambda)&\equiv& \left(\lambda\frac{d}{d\lambda}\right)g^{(-1)}(\lambda)\;,
\eea
with $\lambda=g^2N$, $f(\lambda)$ the cusp anomalous dimension, and $g(\lambda)$ the collinear anomalous dimension.
Expanding in loops $L=$ powers of $\lambda$, it is easy to see that we have 
\be
M_\epsilon^{(L)}=\left(\frac{\mu^2}{-t}\right)^{L\epsilon} f_L(s/t,\epsilon)\;,
\ee
where $f_L(s/t,\epsilon)$ has a pole of order $1/\epsilon^{2L}$.

Alternatively, as found in \cite{Gonzalez:2020tpi}, expanding in $a^L$ (instead of $\lambda^L$), one finds in terms of $r=-s/t$ that
\be
M_\epsilon^{(L)}=\left(\frac{\mu^2}{-t}\right)^{L\epsilon}{\cal F}_L(r,\epsilon).
\ee

Then, at one-loop, the celestial amplitude $\tilde {\cal A}(\{\Delta_i,J_i,z_i,\bar z_i\})$ defined in (\ref{eq:Celestial4pt})
has an $f(r,\bar r)$, from (\ref{eq:f}), of a similar form with the tree level one: We first identify 
$\cB(s,t)=a \left(\frac{\mu^2}{-t}\right)^\epsilon {\cal F}_1(r,\epsilon)$, then we do the integral over $w=-t$ in (\ref{eq:f}) as 
\be
\int_0^\infty \frac{dw}{w} w^{\frac{i\lambda}{2}}\cB(rw,-w)=a\mu^{2\epsilon}{\cal F}_1(r,\epsilon)\int_0^\infty\frac{dw}{w}
w^{i\frac{\lambda}{2}-\epsilon}\;,
\ee
and then note that, because of the $(-t)^{-\epsilon}$ factor, leading to $w^{-\epsilon}$ in the exponent, we shift the argument
of the function $\mathcal{I}(\lambda)$ in the tree celestial amplitude. We also have the extra factor ${\cal F}_1(r,\epsilon)$, for 
a one-loop result
\be
f_{\rm one-loop}(r,\bar r,\epsilon)= 2a g^2 \mu^{2\epsilon} {\cal F}_1(r,\epsilon)
\delta(r-\bar{r}) \Theta(r-1) r^{1+\frac{\Delta}{6}} (r-1)^{\frac{\Delta}{6}} \mathcal{I}(\lambda+2i\epsilon).
\ee

Finally then, the celestial one-loop amplitude is written as an operator acting on the tree amplitude, 
\be
\tilde{\cA}_{\rm one-loop}=a{\cal F}_1(r,\epsilon)(\hat {\cal P})^\epsilon\tilde{\cA}_{\rm tree}\;,
\ee
where the translation operator is 
\be
\hat{\cal P}=\mu^2 r^{\frac{1}{3}}(r-1)^{\frac{1}{3}}\prod_{i<j}(z_{ij}\bar z_{ij})^{-\frac{1}{6}}\exp\left(\frac{i}{2}\sum_{k=1}^4
\frac{\d}{\d\lambda_k}\right).
\ee

This translation operator is related to the translation operator $P_{+,k}$ that shifts the conformal dimension of a single gluon by 
one unit, $\Delta_k\rightarrow \Delta_k+1$, but involves the {\em product}, instead of the sum, over all $k$. 
This $P_{+,k}$ can be also used for the description of IR divergences, but in the case of an IR momentum cut-off for loop integrals, 
as in \cite{Arkani-Hamed:2020gyp}.

Then at $L$-loops, we have similarly 
\be
\cB(s,t)=a^L \left(\frac{\mu^2}{-t}\right)^{L\epsilon}{\cal F}_L(r,\epsilon)\;,\label{BLloop}
\ee
leading to $\mathcal{I}(\lambda)\rightarrow\mathcal{I}(\lambda+2iL\epsilon)$, and finally 
\be
\tilde{\cA}_{\rm L-loop}=a^L {\cal F}_L(r,\epsilon)(\hat {\cal P})^{L\epsilon}\tilde{\cA}_{\rm tree}.
\ee

Then, if we take the actual ${\cal F}_L$ obtained from the BDS formula, we see that the total amplitude is an exponential 
factor, that now acts as an operator, acting on the tree amplitude.

\section{Celestial gravity amplitudes and IR divergences}

For $n$-graviton amplitudes in a gravitational theory, 
similarly we obtain IR divergences expressed as poles in $\epsilon$ for 
$D=4-2\epsilon$ dimensions. 

It was shown in \cite{Naculich:2011ry} that we can separate the amplitude in a hard part $H_n$ that is finite in $\epsilon$, 
times a soft function $S_n$, that contains all the IR divergences,
\be
A_n=S_n\cdot H_n\;,\label{factorgrav}
\ee
where the soft function is 
\be
S_n = \exp \biggl[  {\sigma_n\over\epsilon} \biggr],
\qquad \sigma_n =   {\lambda_{\rm GR} \over 16 \pi^2} 
\sum_{j=1}^n \sum_{i<j}  
s_{ij} \log \left( -s_{ij} \over \mu^2 \right), 
\qquad s_{ij} = (k_i + k_j)^2 \;,
\label{soft}
\ee
where $\lambda_{\rm GR}\equiv (\kappa/2)^2(4\pi e^{-\gamma})^\epsilon$.
Note that, unlike nonabelian gauge theories, like the case of ${\cal N}=4$ SYM treated above, 
there are no jet functions, since we have no collinear singularities in gravity, only in nonabelian gauge theory. As a result, 
the IR divergent factors in the exponent contain only single $1/\epsilon$ poles, not double  $1/\epsilon^2$ poles.

However, note that, since for (massless) gravitons $\sum_i s_{ij}=0$, $S_n$ can be rewritten as a formal sum with $1/\epsilon^2$
poles, more like the nonabelian gauge theory case, 
\be
S_n=\left\{\prod_{i<j}\exp\left[\frac{1}{\epsilon^2}\frac{\lambda_{\rm GR}(-\mu^2)}{16\pi^2}\left(\frac{-s_{ij}}{\mu^2}\right)^{1-\epsilon}\right]
\right\}_{\rm div. }\equiv \tilde S_n|_{\rm div}
\ee

The object in the exponent in $S_n$ can be rewritten as 
\be
\frac{1}{\epsilon^2}\left(\frac{\lambda_{\rm GR}(-\mu^2)}{16\pi^2}\right)\left(\frac{-s_{23}}{\mu^2}\right)^{1-\epsilon}\times \left[
1+\sum_{i<j; (i,j) \neq (2,3)}\left(\frac{s_{ij}}{s_{23}}\right)^{1-\epsilon}\right]\;,\label{exponent}
\ee
in a similar manner with what happened in the case of ${\cal N}=4$ SYM in the previous section.

We want to consider the effect of this factorization of IR divergences on the celestial graviton amplitudes. 

The tree level 4-amplitude (which is the same in all gravity theories, including ${\cal N}=8$ supergravity) 
is written as a square of gauge theory amplitudes, 
\be
A_4^{(0)}=\frac{16i K^2}{stu}\;,
\ee
where $K$ is the same kinematical factor, involving helicities and momenta, as in the nonabelian gauge theory, defined further below
in (\ref{4-tree}). But we saw that the tree-level MHV gluon amplitude had the form  (\ref{eq:4point}), with a constant $\cB(s,t)=g^2$, so 
we infer that the graviton one also has the same form (\ref{eq:4point}), with a constant $\cB(s,t)=(\lambda_{GR}u)^2$ (the effective 
gravitational coupling, up to a number). 

Now we need to consider the factorization of gravity amplitudes (\ref{factorgrav}), and see from it, analogously to the ${\cal N}=4$ 
SYM case reviewed in the previous section, what that implies for the loop amplitudes, and their IR divergences. We have 
to write a form for the function $\cB(s,t)$ as in (\ref{BLloop}), in terms of a standard IR object, $\left(\frac{\mu^2}{-t}\right)^\epsilon$
times a function of $r=-s/t$, which will then get outside the integral over $w=-t$, and thus remain unchanged by the Mellin transform 
of the celestial amplitude.

Before proceeding, let us first express each object in the soft factorization \eqref{factorgrav} in terms of the loop expasion, \emph{i.e.},
\eal{
A_n = \sum_{L=0}^\infty A_n^{(L)} \quad ; \quad 
S_n = 1 + \sum_{L=1}^\infty S_n^{(L)} \quad ; \quad
H_n = A_n^{(0)} + \sum_{L=1}^\infty H_n^{(L)}\;,
}
where $A_n^{(0)}$ is the tree level amplitude and $H_n^{L}$ are all infrared-safe hard factors.  Thus, for instance, at one-loop we have 
\be
\label{1-loop}
A_n^{(1)} = S_n^{(1)} A_n^{(0)} + H_n^{(1)}
\ee

In the absence of mass terms, there will be an effective dimensionless coupling that appears by absorbing the 
required powers of momenta (the only dimensionful quantities). In the case of gravity, that is $\lambda_{\rm GR}s_{ij}$. 
The dimensional transmutation scale 
$\mu$ can only appear in the IR divergent part, and to fix overall dimensions. Then in 
the hard (IR finite)  part of the amplitude $H_n$, we can only have the physical dimension of the 
amplitude, given by the corresponding power of $s$, times a function of $s/t$. More precisely then, in this case, we must have 
\be
H_n=A_{n,{\rm tree}}f(\lambda_{\rm GR} s_{23}; s_{ij}/s_{23})\;,
\ee
or, for $n=4$,
\be
H_4 =A_{4,{\rm tree}}f(\lambda_{\rm GR} t; s/t)\;,
\ee
and the expansion in $\lambda_{\rm GR} t$ gives us the loop expansion. Thus, in particular, at one-loop, we have
\be
H_4^{(1)}=\lambda_{\rm GR} t  f^{(1)}(s/t)A_{4,{\rm tree}}
\ee
and $A_{4,{\rm tree}}$ contains the overall dimension as a power of $s$, times yet another function of the ratio $s/t$. 
In the one-loop amplitude, from \eqref{1-loop} we see that, besides this hard factor there is an extra contribution coming 
from the soft factor $S_4$, which at one-loop is given by just the exponent in (\ref{exponent}), 
\be
\frac{1}{\epsilon^2}\left(\frac{-\mu^2 \lambda_{\rm GR}}{16\pi^2}\right)\left(\frac{-t}{\mu^2}\right)^{1-\epsilon}\left[2+2\left(\frac{s}{t}\right)^{1-
\epsilon}\right].
\ee

Then  the $\cB(s,t)$ coming from the one-loop amplitude is the constant from the tree amplitude, times $(-\mu^2 \lambda_{\rm GR})
\left(\frac{-t}{\mu^2}\right)^{1-\epsilon}$  
times a function of $s/t$, which means that the function $f_{\rm one-loop}(r,\bar r,\epsilon)$ 
contains $(\lambda_{\rm GR}(-\mu^2)) \mu^{2\epsilon}$,  a function of the cross ratio, $\tilde{\cal F}_1(r,\epsilon)$, and the integral
\be
\int_0^\infty \frac{dw}{w} w^{i\frac{\lambda}{2}+1-\epsilon}=\mathcal{I}(\lambda-2i(1-\epsilon)).
\ee

Similarly, again, at $L$ loops we find 
\be
\cB(s,t)\big|_{\rm div}=\frac{1}{L!}(-\mu^2 \lambda_{\rm GR})^L\left(\frac{\mu^2}{-t}\right)^{L(-1+\epsilon)}\tilde 
{\cal F}_L(r,\epsilon)\big|_{\rm div}\;,
\ee
and so 
\be
\label{L-loop-celestial-grav}
\tilde{\cA}_{\rm L-loop}\big|_{\rm div}=(-\mu^2 \lambda_{\rm GR})^L\tilde{\cal F}_L(r,\epsilon)
(\hat {\cal P})^{L(-1+\epsilon)}\tilde {\cA}_{\rm tree}\big|_{\rm div}.
\ee

For the first two terms in the expansion in $\epsilon$ of the $L$-loop amplitude, we do not actually need to know the full hard part
of the gravity amplitude, {\em only the one-loop amplitude} of the corresponding theory, since we have
in any gravity theory \cite{Naculich:2013xa}
\be 
A_n^\Ell (\eps)=
 {1 \over (L-1)!}  \biggl[ {\sigma_n \over \epsilon} \biggr]^{L-1} 
A_n^\One(L\eps)  
+ \cO(1/\eps^{L-2} )\;,
\label{Lloopproponeloop}
\ee
which is an operator acting on the one-loop amplitude $A_n^\One$, but depending on $L\epsilon$ instead of $\epsilon$. 
Thus in this case, knowing ${\cal F}_1(r,\epsilon)$ is enough to calculate the first two terms in the expansion in 
$\epsilon$ of the $L$-loop celestial amplitude. 

Finally then, the IR divergences exponentiate also in the celestial amplitude, now in terms of the operator 
$\hat {\cal P}^{-1+\epsilon}$, and we therefore find also the same soft/hard factorization for the celestial case. 

We note that the IR divergences of graviton celestial amplitudes were considered also in \cite{Arkani-Hamed:2020gyp}, however in the case 
of IR momentum cut-off regularization for loop integrals, $\Lambda_{IR}$. Then the factorization (\ref{factorgrav}), with 
\be
S_n= \exp\left\{-\frac{2G_N}{\pi}\ln \Lambda_{\rm IR}\sum_{i,j} p_i\cdot p_j \ln (p_i\cdot p_j)\right\}\;,
\ee
with $s_{ij}=2p_i\cdot p_j$, leads to a similar one for celestial amplitudes, 
\be
\tilde{\cA}_n=\tilde{\cA}_{\rm soft}\cdot \tilde{\cA}_{\rm hard}\;,
\ee
where 
\bea
\label{softfactor-alternative}
\tilde{\cA}_{\rm hard}&=&\prod_i\int_0^\infty d\omega_i \omega_i^{\Delta_i-1} H_n\cr
\tilde{\cA}_{\rm soft}&=& \exp\left[\frac{2G_N}{\pi}\ln \Lambda_{\rm IR}\sum_{i,j} P_{i}P_{j}|z_{ij}|^2\log |z_{ij}|^2\right].
\eea

This result is consistent with ours, though our methods are different.

\section{IR divergences of celestial  gluon amplitudes: most-subleading-color}

In this section we will consider the IR divergences of celestial gluon amplitudes in a general gauge theory, and we will note that, 
as in the case of usual amplitudes, the most-subleading-color term has properties that are similar
to the ones of gravity, making it easier to write a factorized expression. 

\subsection{Set-up: color and $N$ expansion, factorization}

The full amplitude is expanded in a trace basis $\{T_\lambda\}$ of single and multiple-traces
of generators $T_a$ in the fundamental representation
\be
\cA_n = \sum_\lambda T_\lambda A_{n,\lambda}\;,
\label{basis}
\ee
where the coefficients $A_\lambda$ are the color-ordered amplitudes. The $A_\lambda$ elements are 
organized into a ket $\ket{A}$, according to the standard notation in \cite{Catani:1996jh,Catani:1998bh}. 

For 4-gluon amplitudes ($n=4$), $\lambda=1,...,9$ (six single traces and 
three double traces), however, since only three out of the six single traces are independent, one can choose a 
basis using only three of them. A useful such basis is given by
\bea
T_1 &=&  \Tr(1234) + \Tr(1432),
\qquad\qquad
T_4 =  2 \Tr(13) \Tr(24) , \cr
T_2 &=& \Tr(1243) + \Tr(1342),
\qquad\qquad
T_5=  2 \Tr(14) \Tr(23)  ,
\qquad\qquad 
\label{fourpointbasis}  \\
T_3 &=&  \Tr(1324) + \Tr(1423),
\qquad\qquad
T_6 =  2 \Tr(12) \Tr(34) . \nonumber
\eea

In this basis, the 4-gluon tree level amplitude is written as
\be
|A_4^{(0)}\rangle=-\frac{4iK}{stu}(u, t, s, 0, 0, 0)^T\;,\label{4-tree}
\ee
where $K$ is a factor depending on the helicities and momenta of the external gluons and can 
be found in equation (7.4.42) in \cite{Green:1987sp}. 

At loop level, for the gauge group $SU(N)$ (the relevant case for phenomenology), the amplitudes can be expanded in both 
$L$ and $1/N$, with coefficients $\ket{A^\Ellk}$, 
\be
\ket{A} = \sum_{L=0}^\infty \sum_{k=0}^L {\as^L \over N^k } \ket{A^\Ellk}\,,
\label{expansion}
\ee
where 
\be
a(\mu^2) = {g^2(\mu^2) N \over 8 \pi^2} (4 \pi e^{-\gamma} )^\eps 
\ee 
is the 't Hooft coupling
and $\mu$ is the renormalization scale. Here $k\leq L$ and, anticipating our discussion of the following subsection, notice 
that  for $k=L$ there is no overall power of $N$ multiplying $\ket{A^\Ellk}$. These (infinite) subset of amplitudes are the ones 
known as {\em most-subleading color} gluon amplitudes.

At one-loop, we then only have $\ket{A^{(1,0)}}$ and $\ket{A^{(1,1)}}$. However, unlike the tree level amplitude, these 
are theory-dependent functions. For the case of ${\cal N}=4$ SYM, we have
\be
A_{[1]}^{(1,0)}=2iK I_4^{(1)}(s,t)
\ee
and
\be
A_{1234}^{(1,1)}\equiv A_{[8]}^{(1,1)}=4iK\left[I^{(1)}_4(s,t)+I^{(1)}_4(t,u)+I^{(1)}_4(u,s)\right]\;,
\ee
where the massless scalar box $I^{(1)}_4(s,t)$ has an exact expression to all orders in $\epsilon$, given by 
\bea
I_4^{(1)}(s,t)&=&-i\mu^{2\epsilon}e^{\epsilon \gamma}(4\pi)^{2-\epsilon}I_4(s,t)\;, \cr
I_4(s,t)&=&\frac{2}{st}\frac{ic_\Gamma}{\epsilon^2}\left[t^{-\epsilon} {}_2F_1(-\epsilon,-\epsilon;1-\epsilon;1+t/s)
+s^{-\epsilon}{}_2F_1(-\epsilon,-\epsilon;1-\epsilon;1+s/t)\right]\cr
c_\Gamma&=&\frac{1}{(4\pi)^{2-\epsilon}}\frac{\Gamma(1+\epsilon)\Gamma^2(1-\epsilon)}{\Gamma(1-2\epsilon)}.
\eea

Note that $I_4^{(1)}(s,t)$ can be written as $t^{-2-\epsilon}$ times a function that only depends on the 
cross-ratio $r=-s/t$. The explicit expansion of $I^{(1)}_4(s,t)$ in powers of $\epsilon$ reads
\bea
I_4^{(1)}(s,t)
&=&
\frac{ 1 } { st}
\Bigg\{ 
 \frac{2}{\epsilon^2}\left(\frac{\mu^2}{-s}\right)^\epsilon
+\frac{2}{\epsilon^2}\left(\frac{\mu^2}{-t}\right)^\epsilon
- \log^2 \left( s \over t\right) -\frac{4\pi^2}{3}
+ \cO(\epsilon) 
\Bigg\}
\\
&=& \frac{ 1} { st}
\Bigg\{ 
 \frac{4}{\epsilon^2}
-\frac{2}{\epsilon} \log\left(-s \over \mu^2\right)
-\frac{2}{\epsilon} \log\left(-t \over \mu^2\right)
+ 2 \log\left(-s \over \mu^2\right) \log\left(-t \over \mu^2\right)
-\frac{4\pi^2}{3}
\Bigg\}\,.
\nonumber
\eea

From the first line above we also note that, when Mellin transforming these one-loop expressions, 
the first two terms will simply amount to the change ${\cal I}(\lambda)\rightarrow {\cal I}(\lambda+i\epsilon)$ 
(with $\cal I(\lambda)$ defined in \eqref{eq:deltalambda}) plus constant terms since they are either pure 
numbers or they depend on ratios of Mandelstam invariants only.

For ${\cal N}=1,2$ SYM theories, the full one-loop amplitude is (along with its $A^{(1,0)}$ and $A^{(1,1)}$ components), 
to the best of our knowledge, not yet available in the literature. However, the general form of the amplitude being 
of the type $t^{-2-\epsilon}$ times a function of $r=-s/t$ is still valid here. Moreover, it is not only true at the 
one-loop level but at {\em all-loop} orders.

In order to see this, let us consider the following argument, which is actually even simpler than the one used 
for the case of gravity in the previous section. The theories we are considering here have no scales at 
all since the couplings are dimensionless. Therefore, in general, we must have that the IR-finite 
part (\emph{i.e.}, the hard factor) 
consists in an overall power of $s_{ij}$, 
(maybe times a power of $s_{ij}/\mu^2$ for the anomalous dimension, if it is non-zero, although this is
 usually taken care of by renormalization and by the factoring of IR divergences), 
times another function depending purely on the ratios $s_{ij}/s_{kl}$ and of the coupling, that is
\be
\left|H\left(\frac{s_{ij}}{\mu^2},a(\mu^2),\epsilon\right)\right\rangle=A_{n[i]}^{(0)}\left(s_{12},\frac{s_{ij}}{s_{12}}\right)
f_{[i]} \left(a(\mu^2), \frac{s_{ij}}{s_{12}}\right)
\;,
\ee
where the $[i]$ index on the right refers to the $i$-th component of the ket $\ket{H}$. For the $n=4$ case, this reads 
\be
\left|H\left(\frac{s_{ij}}{\mu^2},a(\mu^2),\epsilon\right)\right\rangle=A_{4[i]}^{(0)}(t, r) f_{[i]}(a(\mu^2),r)\;,
\ee

Note that the full loop expansion is obtained from the factor $f(a(\mu^2),r)$ which is just a function of the ratio $r=-s/t$.

In the above relations we have used a notation for the IR finite part that previews an important fact about general
color-ordered gluon amplitudes: they can be factorized into the product of three objects: a jet function $J$, 
a soft factor $\bS$, and a hard IR-safe factor $H$. The jet factor $J$ is just a scalar function (a number) 
while the soft function $\bS$ is a matrix-valued object (defined by the soft anomalous dimension matrix 
$\bGam$ below). Both jet and soft factors are IR divergent, thus describing the large distance behaviour, 
while the hard function $H$, being IR-finite (as $\epsilon
\rightarrow 0$), describes short distance physics. In the \emph{ket} notation for the amplitudes, the 
factorization reads \cite{Sterman:2002qn,Aybat:2006mz}
\be
\left|   A \left(\mommu,  \as, \ep\right) \right> 
= 
 J \left(\as, \eps \right) \, 
{\bS} \left( \mommu,  \as, \eps\right) 
\left | H \left( \mommu, \as, \eps \right) \right>.
\label{YMfactor}
\ee

The soft function, for any gauge theory, is generically 
\be
{\bS} \left( \mommu, \as, \eps \right) 
\,=\,
{\rm P}~{\rm exp}\left[
\, -\; 
\frac{1}{2}\int_{0}^{\mu^2} 
\frac{d\tilde{\mu}^2}{\tilde{\mu}^2}
\bGam \left( \mommu,
         \bas \left(\frac{\mu^2}{\tilde{\mu}^2}, \as, \eps  \right)
       \right) 
\right]\;,
\label{YMsoft}
\ee
where the central object $\bGam$ is known as the soft anomalous dimension matrix.\footnote{The exact relation 
between $\bar a$, $a$, and the dimensional regulator $\epsilon$ is rather involved (see, \emph{e.g.}, Eq.(7) in 
\cite{Sterman:2002qn}), but it is not relevant for our purposes here. See also \cite{Magnea:2000ep} for earlier 
work.} The loop expansion of $\bGam$,
\be
\label{soft-matrix}
\bGam= \sum_{L=1}^\infty a(\mu^2)^L \bGam^{(L)}\;,
\ee
is fully known (for massless theories) up to three loops \cite{Almelid:2015jia}, while four-loop corrections 
have recently appeared in the literature \cite{Becher:2019avh}. The one-loop matrix is 
\be
\bGam^\One = \frac{1}{N}  
\sum_{j=1}^n \sum_{i<j}  
\bT_i \cdot \bT_j \log \left( {\mu^2 \over -s_{ij}  } \right) \;,
\label{oneloopanom}
\ee 
where $\bT_i$ is the color charge operator for the $i$-th gluon corresponding to the SU$(N)$ generators in the 
adjoint representation. Here $\bT_i \cdot \bT_j\equiv T_i^a T_j^a$, $\bT_i^2=C_i$, with $C_i$ being the 
quadratic Casimir where, in the adjoint representation, $C_i=N$. The matrix element of $\bT_i$ in between states 
characterized by gluon adjoint indices $b_i,c_i$ is 
\be
\langle c_1,...,c_i,...c_m|\bT_i^c|b_1,...,b_i,...,b_m\rangle=\delta_{c_1b_1}...T^c_{c_ib_i}...\delta_{c_mb_m}\;,
\ee
where $T^c_{c_ib_i}=if_{c_icb_i}$. Another way of characterizing these operators is by writing how they act upon the generators
\be
\bT_i^a T_{b_i}=i{f^a}_{b_ic_i}T_{c_i}=[T_a,T_{b_i}]\;,
\ee
which defines the action on the color basis. 

Given this action of $\bT_i$, in terms of the color-ordered expansion (\ref{basis}),
the $L$-loop anomalous dimension matrix $\bGam^\Ell$ acts on a given element $T_\lam$ of the trace basis
(\ref{basis}) to yield a linear combination
\be
\bGam^\Ell  T_\lambda =
\sum_\kappa T_\kappa  \bGam^\Ell_{\kappa \lambda}\;,
\label{action}
\ee
and it is the matrix $\bGam^\Ell_{\kappa \lambda}$
that then acts on the hard ket $\ket{H}$. 

The fact that the two-loop contribution to the anomalous dimension matrix is proportional to the one obtained at 
one-loop, together with arguments based on collinear limits and symmetry considerations regarding rescalings 
of the momenta of the external states, led to a fascinating conjecture coined {\em the dipole conjecture}, valid 
for any gauge theory, including QCD. This stated that all higher loop corrections in the soft anomalous dimension 
matrix $\bGam$ were proportional to the one-loop matrix $\bGam^{(1)}$ \cite{Becher:2009cu,Gardi:2009qi,Becher:2009qa}. 
However, as argued in \cite{Caron-Huot:2013fea} and later shown by an explicit 3-loop calculation \cite{Almelid:2015jia}, 
the dipole conjecture fails starting at three-loop order. \footnote{This is true for a generic gauge theory, in particular, for the
phenomenologically important case of QCD.}

Within the dipole approximation, it is easy to see from \eqref{oneloopanom} that $[\bGam(\mu_1),\bGam(\mu_2)]=0$, 
for any two different scales $\mu_1\neq \mu_2$. This makes the path ordering operator in \eqref{YMsoft} imaterial 
allowing us to directly compute the integral, yielding
\be
\bS \left( \mommu, \as, \epsilon \right) 
 = \exp \left[ \sum_{L=1}^\infty 
{\as^L  \over  2 L \epsilon }
{\bGam^\Ell} 
\left(1+ {\cO} \left(\frac{\as}{\epsilon}\right)\right)
\right].
\label{integrated}
\ee

{\em Note however that in this case, we still need to calculate all of the $\bGam^\Ell$ in order to have the full IR behaviour 
in the soft function $\bS$}. We would indeed have an exponentiation due to the resummations, but in the exponent we still 
have to sum over loops. Moreover, one would also need to calculate the jet function $J$. 

In the rest of this section we will focus on the effect of the dipole structure and in Section~\ref{sec:nondipole} we will 
explore corrections from non-dipole terms. 

Under the dipole conjecture, {\em and considering only the one-loop anomalous dimension contribution in the 
exponent in} (\ref{integrated}) to all orders in $N$, we obtain
\be
\bS \left( \mommu, \as, \epsilon \right) =\exp\left[\frac{a}{2\epsilon}\bGam^\One\right]
 = \exp \left[\frac{a}{2\epsilon^2}\frac{1}{N}  \sum_{j=1}^n \sum_{i<j}  \bT_i \cdot \bT_j 
\left\{\left(\frac{-s_{ij}}{\mu^2}\right)^{-\epsilon}-1\right\}\right]\;,\label{dipoleSone}
\ee
with the understanding that, in the second equality, only the divergent and finite terms in the $\epsilon$ 
expansion are to be kept. Note that, due to color conservation $\sum_{i=1}^n \bT_i=0$,
\be
\sum_{j=1}^n \sum_{i<j}  \bT_i \cdot \bT_j = -\frac{1}{2}\sum_{j=1}^n \bT_j^2=-\frac{1}{2} n N\one \,,
\ee
the second term in the exponent of (\ref{dipoleSone}) commutes with the first. Therefore, only 
the action of the first term upon the hard amplitude $\ket{H}$ is non-trivial. In the four gluon case, $\sum_{i<j}\bT_i\cdot \bT_j
\left(\frac{-s_{ij}}{\mu^2}\right)^{-\epsilon}$ gives
\be
\left(\frac{-t}{\mu^2}\right)^{-\epsilon}\left[\bT_2\cdot \bT_3+\bT_1\cdot\bT_4+\left(-\frac{s}{t}\right)^{-\epsilon}\left(\bT_1\cdot \bT_2
+\bT_3\cdot \bT_4\right)+\left(-\frac{u}{t}\right)^{-\epsilon}\left(\bT_1\cdot \bT_3+\bT_2\cdot \bT_4\right)\right]\;,
\ee
from where we see that the square bracket is a function of $r=-s/t$ only, while the prefactor is the 
standard one that already appeared in ${\cal N}=4$ SYM and in gravity. 

So in this case, yet again, when calculating the celestial amplitude,
$f_{\rm one-loop}(r,\bar r,\epsilon)$ {\em for the first term in the exponent of $\bS$}
contains $a\mu^{2\epsilon}$, a function ${\cal F}_1(r,\epsilon)$, 
that gets outside of the integral over $w=-t$, and the same ${\cal I}(\lambda+2i\epsilon)$, since we have 
\be
\int_0^\infty \frac{dw}{w} w^{\frac{i\lambda}{2}}\cB(rw,-w)=a\mu^{2\epsilon}{\cal F}_1(r,\epsilon)\int_0^\infty\frac{dw}{w}
w^{i\frac{\lambda}{2}-\epsilon}\;,
\ee
and so involves the same translation operator $\hat P$ as before. Its action is now not on the tree amplitude, but rather on the 
hard amplitude $|H\rangle$,  which is a function of $r$ (times the trivial dependence on $t$ given by its dimension).
The second term is just a constant. Then, at $L$ loops, we obtain 
\be
\cB(s,t)=\frac{a^L}{L!} \left[\left(\frac{\mu^2}{-t}\right)^{\epsilon}-(...)1\right]^L{\cal F}_L(r,\epsilon)\;,
\ee
yielding $\mathcal{I}(\lambda)\rightarrow\mathcal{I}(\lambda+2iL\epsilon)$ in the leading term, and so 
\be
|\tilde{\cA}_{\rm L-loop}\rangle=a^L {\cal F}_L(r,\epsilon)\left[(\hat {\cal P})^{\epsilon}-(...)1\right]^L|H\rangle.
\ee

Finally, note that 
\be
\label{1-loop-gamma}
\bGam^\One=2\begin{pmatrix}\a&0\\0&\delta\end{pmatrix}+\frac{2}{N}\begin{pmatrix} 0&\beta\\ \gamma& 0\end{pmatrix}\;,
\ee
and $\a,\delta$ are written in terms of ${\cal S}$, ${\cal T}$, ${\cal U}$ themselves, or their {\em sums}, where 
\be
{\cal S}=\log\left( \frac{-s}{Q^2}\right)\;,\;\;
{\cal T}=\log\left(\frac{-t}{Q^2}\right)\;,\;\;
{\cal U}=\log\left(\frac{-u}{Q^2}\right)\;,
\ee
whereas $\b,\gamma$ are written in terms of {\em differences } of the same, so are functions of 
ratios of $s,t,u$. This implies that the leading (in $N$) part of $\Gamma^{(1)}$ depends independently of $s,t,u$, whereas the subleading 
(in $1/N$: the part with $\beta$ and $\gamma$) part, $\Gamma^{(1)}_{\rm sub}$, depends only on the ratios.

\subsection{IR divergences for the most-subleading-color amplitudes}

We have seen that we can write a formula for the IR divergence of the celestial gluon amplitudes using 
the dipole conjecture, but considering only the one-loop term in the exponent. 

However, and this is the crucial point, if we focus on the (still infinite set of) most-subleading-color 
amplitudes at all loops, $|A^{(L,L)}\rangle$, which contain only terms 
of order $N^0$ (the others contain positive powers of $N$) we can {\em completely 
determine} the IR divergence \cite{Naculich:2013xa}. Indeed, while in general $\bGam^\Ell_{\kappa \lambda}$ possesses all 
powers of $N$, from ${\cal O}(1)$ to ${\cal O}(1/N^L)$ (so that, when multiplied with $a^L\propto N^L$, we get powers $N^L$
through 1), the dipole conjecture says that $\bGam^\Ell_{\kappa \lambda}$ is proportional to $\bGam^\One_{\kappa \lambda}$, 
which contains only ${\cal O}(1)$ and ${\cal O}(1/N)$. Thus, when multiplied with $a^L$, it gives terms of ${\cal O}(N^L)$ and 
${\cal O}(N^{L-1})$, so does not contribute to the most-subleading-color amplitude. Therefore, we obtain that 
{\em the most-subleading-color part of the amplitude has an one-loop (though exponentiated) exact IR divergence }, 
\be
\ket{A_n} \bigg|_{\rm most-subleading-color}
 = \exp \left[ \frac{\as}{2\epsilon} 
 \bGam^\One_{\rm sub}   \right]
\ket{H_n(\eps)}\bigg|_{\rm most-subleading-color}
\label{YMoneloopexact}
\ee

For the 4-point function, this becomes
\be
\ket{A_4} \bigg|_{\rm most-subleading-color}
=  \exp \left[ 
{\as \over N \epsilon} 
\left( \begin{array}{cc} 0 & \beta \\ \gamma & 0 \end{array} \right)
\right]
\ket{H_4(\eps)} \bigg|_{\rm most-subleading-color} \,,\label{YM4oneloopexact}
\ee
where
\be
\beta = 
\left( 
\begin{array}{ccc}
0 & -2Y & 2X \\ 
2Z &  0 &- 2X \\ 
-2Z&  2Y & 0  
\end{array}
\right) ,
\qquad
\gamma =
\left( 
\begin{array}{ccc}
0 & -X & Y \\
 X & 0 & - Z \\
-Y & Z & 0
\end{array}
\right)
\qquad
\ee
and
\be
X =\log \left(t \over u\right), \qquad
Y =\log \left(u \over s\right), \qquad
Z =\log \left(s \over t\right).
\label{XYZ}
\ee

For the celestial amplitude, we note that these functions are all functions of $r=-s/t$, so in the $\cB(s,t)$ integral over $w=-t$, 
they become mere constants that we can take out of the integral. Therefore the IR divergence is a simple 
constant factor leaving only the integral of the tree amplitude, yielding $\cal {I}(\lambda)$. Therefore, 
the formula (\ref{YMoneloopexact}) continues to hold for the celestial amplitude.

Moreover, for the first two leading divergencies for each most-subleading amplitude $A^{(L,L)}$, for $L=2l$ (even) or $L=2l+1$ (odd), 
we have the relations (first conjectured for ${\cal N}=4$ SYM in \cite{Naculich:2008ys} and then 
proven in \cite{Naculich:2009cv}, if and when the dipole conjecture holds, for a general gauge theory in \cite{Naculich:2013xa}):
\bea
\left(
\begin{array}{c}
A_1^\Tete (\epsilon) \\
A_2^\Tete (\epsilon) \\
A_3^\Tete (\epsilon) \\
\end{array}
\right)
&=&
  { 2  (2X^2 + 2Y^2 + 2Z^2)^{\ell-1} \over
 (2\ell-1)! \epsilon^{2\ell-1} }
A_4^\Oneone ((2\ell)\epsilon) 
\left(\begin{array}{c} X-Y \\ Z-X\\ Y-Z\\ \end{array} \right)
+ \cO(1/\eps^{2 \ell-2}) \,,\cr
&&
\qquad\label{evenLYM}
\\
\left(
\begin{array}{c}
A_4^\Teptep (\epsilon)  \\
A_5^\Teptep (\epsilon) \\
A_6^\Teptep (\epsilon) \\
\end{array}
\right)
&=&
{(2X^2 + 2Y^2 + 2Z^2)^{\ell} \over
(2\ell)!  \epsilon^{2\ell}}
A_4^\Oneone ((2\ell+1)\epsilon) 
\ones 
+ \cO(1/\eps^{2 \ell-1}) \,.\cr
&&\label{oddLYM}
\eea

Since, as we saw, $X,Y,Z$ are functions of only ratios of Mandelstam variables, and get outside the $w=-t$ 
integration in the celestial amplitude, equations (\ref{YM4oneloopexact}), 
(\ref{oddLYM}) and (\ref{evenLYM}) for 4-graviton amplitudes are also valid in celestial amplitudes.

Here we should note that the behaviour of (\ref{oddLYM}) and (\ref{evenLYM}) for the most-subleading amplitude 
is similar to the gravity relation (\ref{Lloopproponeloop}). In fact, the relation between the two can be made more precise, 
as was done in \cite{Naculich:2013xa} for the usual amplitudes:
\bea
&&M_4^{(2\ell+1)} (\eps) 
= \left( - {\lambda\over 8 \pi^2 } \right)^{2\ell+1}
 { (sY-tX)^{2\ell} \over (2X^2 + 2Y^2 + 2Z^2)^\ell }  
\quad  { A^\Ellodd  (\eps) \over (A_1^\Zero/u) } 
+\cO \left( \frac{1}{\epsilon^{2\ell-1}}\right)\nn
\\
&=& \left( - {\lambda\over 8 \pi^2 } \right)^{2\ell+1}
\left[\frac{(s\log s +t\log t+u\log u)^2}{2(\log^2(t/u)+\log^2(u/s)+\log^2(s/t))}\right]^\ell
{ A^\Ellodd  (\eps) \over (A_1^\Zero/u) } 
+\cO \left( \frac{1}{\epsilon^{2\ell-1}}\right)
\cr
&&\label{oddlooprelation}
\eea
for odd $L$ and 
\bea
M_4^{(2\ell)} (\eps) 
&=& \left(  {\lambda\over 8 \pi^2 } \right)^{2\ell}
 { (sY-tX)^{2\ell-1} \over (2X^2 + 2Y^2 + 2Z^2)^{\ell-1} }  
\quad { A_1^\Elleven(\eps)  \over 2 (X-Y) (A_1^\Zero/u) }
+\cO \left( \frac{1}{\epsilon^{2\ell-2}}\right)\cr
&=& \left(  {\lambda\over 8 \pi^2 } \right)^{2\ell}
 { (sY-tX)^{2\ell-1} \over (2X^2 + 2Y^2 + 2Z^2)^{\ell-1} } 
{\left(  A_1^\Elleven (\eps) - A_2^\Elleven (\eps) \right)
\over 6X (A_1^\Zero/u) }
+\cO \left( \frac{1}{\epsilon^{2\ell-2}}\right)\cr
&&
\label{evenlooprelation}
\eea
for even $L$ (here $A_1^{(0)}/u=-4iK/(stu)$ is symmetric in $s,t,u$).

Now again $X,Y,Z$ are only functions of the ratios of $s,t,u$, however $s,t,u$ appear explicitly in the relations. We could 
note that, considering the form on the first line(s) of the equation(s), we can take outside an $u^L$ factor in both cases ($L=2l+1$ 
and $L=2l$), obtaining the modified coupling factor $(\lambda t)^L$ in front, and then only ratios of $s,t,u$, and the ratio of the 
amplitude to the tree ampliture, $A^{(L,L)}/A_1^{(0)}$, just like on the left-hand side. Therefore, when going to the celestial 
amplitudes, the relation is the same, except for the extra $w^L=(-t)^L$ in the integral over $w$ giving ${\cal I}(\lambda)$, 
shifting its argument, which, as we saw, means an extra translation operator $\hat{\cal P}$ for each $-t$ factor. 
So the $(\lambda t)^L$ transforms into $(\lambda \hat{\cal P})^L$, but otherwise the relation is unchanged.

For loop number $L=0,1,2$, the relation between the most-subleading 4-point amplitude divided by the tree amplitude 
and gravity 4-point amplitudes divided by their tree amplitudes is actually exact, as shown in \cite{Naculich:2008ys} for 
${\cal N}=4$ SYM vs. ${\cal N}=8$ supergravity and in \cite{Naculich:2011my} at any ${\cal N}$:
\be
\left(\sqrt{2}\lambda_{\rm SYM}\right)^L M^{(L)}_{{\rm SG},{\cal N}+4}(s,t)=\frac{1}{3}\left[\left(\lambda_{\rm SG}u\right)^L
M_{{\rm SYM},{\cal N}}^{(L,L)}(s,t)+{\rm cyclic}\;\;{\rm perms}\;\;{\rm of}\;\; s, t, u\right]\;,
\ee
where $\lambda_{\rm SYM}=g^2N$ and $\lambda_{\rm SG}=(\kappa/2)^2$. Again the same observation holds: we can take a
factor of $w^L=(-t)^L$ common, so as to have $(\lambda_{\rm SG} w)^L$ and the rest just ratios of $s, t, u$ and $M_{\rm SYM}$, 
and then we just shift the integration over $w$, so shift the argument of ${\cal I}(\lambda)$, or act with the translation operator
$\hat {\cal P}^L$, giving $(\lambda \hat{\cal P})^L$.

For the 5-point functions, there are also some relations. At one-loop, we have the exact relation (found for ${\cal N}=4$ 
SYM in \cite{Naculich:2011fw} and proven for general ${\cal N}$ in \cite{Naculich:2011my}) between ratios of loop amplitudes and 
tree amplitudes,
\be
M_{{\cal N}+4}^{(1)}(1,2,3,4,5)=\frac{1}{20}\left(\frac{\kappa}{2}\right)^5\sum_{S_5}i\b_{12345}A^{(1,1)}_{12;345,{\cal N}}\;,
\ee
where $\b_{12345}$ are one-loop numerators defined in \cite{Carrasco:2011mn}. We also have a relation between tree amplitudes
involving the same numerators, found in \cite{Nastase:2010xa} for ${\cal N}=4$ and used at general ${\cal N}$ in 
\cite{Naculich:2011my},
\be
{\cal M}_{{\cal N}+4}^{(0)}(i,j,a,b,c)s_{ij}=\sum_{\sigma\in S_3}\b_{i,j,\sigma(a),\sigma(b),\sigma(c)}A^{(0)}_{i,j,\sigma(a),\sigma(b),\sigma(c)}.
\ee

Then, in the case of 5-point functions, going to the celestial amplitudes would result in more complicated relations between the two 
sides, because of the numerators $\b$, so we will not describe them.

\section{The soft factor and the Knizhnik-Zamolodchikov equation}
\label{sec:KZ}
In this section we consider an important relation of the celestial soft factor to the Knizhnik-Zamolodchikov equation.
We start by reviewing the celestial analog of the soft factorization in nonabelian gauge theory, following 
\cite{Gonzalez:2021dxw} and using their notation. 
In dimensional regularization, the all-loop $n$-gluon amplitude in momentum space takes the 
factorized form (the jet function is now absorbed into the soft factor)
\be
\cM_{n}(\{p_i\}) = \bZ(\alpha(\mu^2),\{s_{ij}\})\cM_{\rm hard} (\{s_{ij}\})\;,
\ee
with the soft factor (previously split into the -matrix- soft function {\bf S} and the -scalar- jet function $J$)
\be
\label{Z}
 \bZ(\alpha(\mu^2),\{s_{ij}\}) = \cP \exp\left\{-\frac{1}{2}\int_0^{\mu^2} \frac{d\lambda^2}{\lambda^2}{\bf \Gamma}_n 
 (\alpha(\lambda^2),\{s_{ij}\})\right\}\;,
\ee
where $\a=2 e^{-\epsilon \gamma_E}g^2/(4\pi)^{2-\epsilon}$ is the YM coupling and  
the (total, i.e. {\em all-loop}) soft anomalous dimension matrix ${\bf \Gamma}_n$ is given by
\begin{multline}
\label{fullGamma}
{\bf \Gamma}_n(\alpha,\left\{s_{ij}\right\} )= -\frac{1}{4N} \gamma(\alpha) \sum^{n}_{i=1} \sum^{n}_{j \neq i} {\bf T}_{i}
\cdot {\bf T}_{j} \log\left( -\frac{s_{ij}}{\lambda^2} \right) -\sum^{n}_{i=1} \gamma_{J_i} (\alpha)\\
+ {\bf \Delta}_n(\alpha, \left\{ \rho_{ijkl} \right\} )+\Delta {\bf \Gamma}_n(\alpha(\lambda^2),\{s_{ij}\})\;,
\end{multline}
where $\gamma(\a)$ is the cusp anomalous dimension.

The first term is the dipole term, described in detail in the previous section (see (\ref{oneloopanom} for its one-loop part, the 
further loops correct just its coefficient, giving the cusp anomalous dimension $\gamma(\a)$ in front), and 
which accounts for pairwise interactions only. The second term is the sum 
over collinear anomalous dimensions of each of the external gluons (jet function contributions in the notation from the previous 
section). The non-dipole contributions ${\bf \Delta}_n$ and $\Delta {\bf \Gamma}_n$ 
make their appearance starting at three \cite{Almelid:2015jia} and four loops \cite{Becher:2019avh} respectively, and were ignored until now.

Using the momentum parametrization \eqref{p}, the Mandelstam variables read $s_{ij}=\eta_{i}\eta_j\omega_i\omega_j |z_{ij}|^2$. 
From here and the color conservation condition 
\be
\label{color-cons}
\sum_{i=1}^n {\bf T}_i =0\;,
\ee
we see that $\bGam_{n}$ reads (we have denoted $\hat\gamma(\a)=\gamma(\a)/C_A$, where $C_A$ is the Casimir in the 
adjoint representation, equal to $N$ here, and $C_k$ refers to the Casimir of gluon $k$, obtained from 
$\sum_{i<j}{\bf T}_i\cdot{\bf T}_j=-\sum_i C_i$ for external gluons)
\be
\label{Gprop}
\bGam_{n}(\alpha,\left\{s_{ij}\right\})= \bGam_{n}\left(\alpha, \{\mu^2|z_{ij}|^2\}\right)+\frac{1}{2}\hat{\gamma}(\alpha) 
\sum^n_{k=1} C_k \log\left(\tfrac{\omega_k}{\mu}\right)\,.
\ee

This separation of $\bGam_n$ into a part logarithmically dependent on the energies $\omega_k$, plus the same function 
evaluated on the celestial data $|z_{ij}|$, allows us to write 
\be
\label{Zprop}
\bZ\left(\alpha, \{s_{ij}\}\right)= \bZ\left(\alpha, \{\mu^2|z_{ij}|^2\}\right)\prod^{n}_{k=1} \left(\tfrac{\omega_k}{\mu}\right) 
^{-\frac{1}{2}\hat{K} C_{k}}\,, 
\ee
where $\hat K(\a)=\frac{1}{2}\int_0^{\mu^2}\frac{d\lambda^2}{\lambda^2}\hat \gamma(\a(\lambda^2))$, 
and therefore, from \eqref{eq:4}, we immediately see that the full $n$-gluon celestial amplitude becomes \cite{Gonzalez:2021dxw}
\be
\tilde{\cM}_{n}(\left\{\Delta_{i},z_i,\bar{z}_i\right\})= \, \bZ\left(\alpha, \{\mu^2|z_{ij}|^2\}\right) \tilde{\cM}_{\rm hard}
\left(\{\Delta_{i}-\tfrac{1}{2}C_i\hat{K} ,z_{i},\bar{z}_i\}\right)\,,
\ee
\emph{i.e.}, the momentum space soft factorization persists after Mellin transforming the amplitudes into celestial correlators. 
Moreover, the {\em full effect, at all loops and all $N$ contributions} (to compare with the previous section, there we considered
only the most subleading component, as well as the one loop in all $N$ components)
of the infrared divergences onto the hard part of the amplitude is a renormalization of the 
conformal dimensions from their bare values $\Delta_i$ to the renormalized ones $\Delta_i-\tfrac{1}{2}C_i \hat K$.
When $\epsilon\rightarrow 0$, $\hat K$ diverges, giving an infinite shift, like it was explicitly found in the case of ${\cal N}=4$
SYM. 

Here we will only concentrate on the contribution from the dipole term in $\bGam_n$, leaving the effects of the 
contributions ${\bf \Delta}_n$ and $\Delta \bGam_n$ for future work, as we did in the previous section. 
In this case, it is easy to see that, for two 
arbitrary scales $\lambda_1\neq \lambda_2$, the commutator $[{\bf \Gamma}_n(\lambda_1^2),{\bf \Gamma}_n
(\lambda_2^2)]$ vanishes, thus making the path-ordering operator $\cP$ in \eqref{Z} immaterial, as explained in the 
previous section. This allows us to write $\bZ$ as
\eal{
\bZ &= \cJ^n \exp\left\{ \frac{K}{4N}\sum^{n}_{i=1} \sum^{n}_{j \neq i} {\bf T}_{i}\cdot {\bf T}_{j} \log\left(|z_{ij}|^2\right)\right\}\;,
}
where we have used color conservation $\sum_{i=1}^n {\bf T}_{i} = 0$ and the definitions
\eal{
K=\int_0^{\mu^2} \frac{d\lambda^2}{\lambda^2}\gamma(\alpha) \quad ; \quad \log\left(\cJ(\alpha,\epsilon)\right)
= \int_0^{\mu^2} \frac{d\lambda^2}{\lambda^2} \left(\frac{N}{4} \gamma(\alpha)\log\left(\frac{-\lambda^2}{\mu^2}\right)
+\gamma_J(\alpha)\right).
}

We have also assumed that that all external states are of the same species (\emph{e.g.}, gluons), that is, 
$\gamma_{J_i}\equiv \gamma_J$, thus
$$
\sum_{i=1}^n \gamma_{J_i}(\alpha)= n \gamma_J(\alpha).
$$

Writing then the infrared factor as the scalar jet function $\cJ^n$ and the soft function $e^{\bA}$, 
\be
\label{Amatrix}
\bZ= \cJ^n e^{\bA} \quad ; \quad \bA = \frac{K}{4N}\sum^{n}_{i=1} \sum^{n}_{j \neq i} {\bf T}_{i}\cdot {\bf T}_{j} \log\left(|z_{ij}|^2\right)\;,
\ee
and taking the derivative of $\bZ$ with respect to any of the positions $z_k$ on the celestial sphere, we have 
\be
\label{BCHexpansion}
\frac{\partial \bZ}{\partial z_k} = \left(\partial_k \bA+\frac{1}{2!}\left[\bA,\partial_k \bA \right]+\frac{1}{3!}
\left[\bA,\left[\bA,\partial_k \bA\right]\right]+\frac{1}{4!}[\bA,\left[\bA,\left[\bA,\partial_k \bA\right]\right]+\cdots\right)\bZ\;,
\ee
where $\partial_k=\partial/\partial z_k$. The first term is
\be
\partial_k \bA = \frac{K}{2N}\sum_{i\neq k} \frac{\bT_k \cdot \bT_i}{z_k-z_i}.
\ee

Thus, if the commutator expansion above can be thought of as a perturbative expansion in a small parameter with 
$\partial_k \bA$ being its leading term, the soft celestial factor $\bZ\left(\alpha, \{\mu^2|z_{ij}|^2\}\right)$ indeed 
satisfies the Knizhnik-Zamolodchikov (KZ) equation in that limit, \emph{i.e.},
\eal{
\left(\frac{\partial}{\partial z_k}-\frac{K}{2N}\sum_{i\neq k} \frac{\bT_k \cdot \bT_i}{z_k-z_i}\right)\bZ = 0.
}

From here, one would identify the infrared divergent factor $K(\epsilon)$ and $N$ with the level $\kappa$ and the Coxeter number $g$ as
\eal{
\frac{K(\epsilon)}{2N} \leftrightarrow -\frac{1}{\kappa+g}.
}

For the particular case of the most-subleading four-gluon amplitude we will see that, at leading order in the Regge limit, 
one indeed obtains the Knizhnik-Zamolodchikov equation. The most-subleading amplitude has the form
\be
\label{Msoft}
\ket{A_n} \bigg|_{\rm most-subleading-color}
 = e^{\bM}
\ket{H_n(\eps)}\bigg|_{\rm most-subleading-color}\;,
\ee
where the matrix $\bM$, for the 4-gluon case at most-subleading order, is given by
\be
\bM = \frac{a(\mu^2)}{\epsilon N}
\begin{pmatrix}
0& \beta \\
\gamma & 0
\end{pmatrix}\;,
\ee
with 
\eal{
\beta = 
\begin{pmatrix}
0 & -2 \log(\tfrac{r-1}{r}) & -2 \log(r-1)\\
2 \log(r) & 0 & 2 \log(r-1)\\
-2 \log(r) & 2 \log(\tfrac{r-1}{r}) & 0
\end{pmatrix} \\
{}\\
\gamma = 
\begin{pmatrix}
0 & -\log(r-1) & \log(\tfrac{r-1}{r})\\
 \log(r-1) & 0 &  -\log(r)\\
-\log(\tfrac{r-1}{r}) & \log(r)& 0
\end{pmatrix}.
}

Here we have changed our momentum conventions, in order to perform the Regge limit in the physical region $(r>1)$. 
Notice that, unlike the general gluon amplitude, the dependence on the $z_k$ coordinates is through the cross-ratio $r$ only. 
Thus, for large $r$, we have,   
\eal{
\beta =  \log(r)
\begin{pmatrix}
0 & 0 & -2\\
2 & 0 & 2\\
-2 & 0 & 0
\end{pmatrix} + \mathcal{O}(r^{-1})\;,
\qquad 
\gamma = \log(r)
\begin{pmatrix}
0 & -1 & 0\\
1 & 0 & -1\\
0 & 1 & 0
\end{pmatrix}+ \mathcal{O}(r^{-1}).
}

Therefore, at leading order in the large $r$ expansion all commutator terms in \eqref{BCHexpansion} vanish, yielding 
the KZ equation in the Regge limit $r\to \infty$.

To conclude, note that the dipole contribution $\bA$ (for finite $N$) can be written as
\eal{
\bA = \bA^{(0)}+\frac{1}{N}\bA^{(1)}.
}

Using this and the fact that $\bA^{(0)}$ is a diagonal matrix (see, \emph{e.g.}, the four-point case in \eqref{1-loop-gamma}) 
the commutator expansion in \eqref{BCHexpansion} organizes into a $1/N$ expansion of the form
$$
\frac{\partial \bZ}{\partial z_k} = \left(\partial_k \bA^{(0)}+\frac{1}{N}\left(\partial_k \bA^{(1)} + [\bA^{(1)},\partial_k 
\bA^{(0)}]+[\bA^{(0)},\partial_k \bA^{(1)}]\cdots \right)+\mathcal{O}(1/N^2)\right)\bZ\,,
$$
thus, in the large $N$ limit, we again see the appearance of the KZ equation. However, since only the diagonal part of
$$
\sum_{i\neq k} \frac{\bT_k \cdot \bT_i}{z_k-z_i}
$$
appears at leading order, the CFT describing the soft celestial part seems to be a theory of non-interacting colored 
scalar fields, as previously hinted in \cite{Magnea:2021fvy,Gonzalez:2021dxw}.

We would like to close this section by mentioning that the appearance of the KZ equation also emerges in 
other kinematical limits, and for the general gluon amplitude, {\em i.e.}, not only for the most-subleading-color 
amplitude (within the dipole aproximation). For instance, in collinear limits such as $z_{12}=z_{34}=\delta\to 0$, 
one also obtains that the commutators in \eqref{BCHexpansion} are subleading with respect to the first term in 
the expansion in powers of $\delta$, although again the nonabelian structure seems to be too simple, similar to 
the large $N$ case. It would certainly be interesting to explore these limits in more general situations, such as 
in higher-point amplitudes and in other multi-collinear regimes. 

Finally, note that in \cite{DelDuca:2011ae}, the authors obtained the Regge limit of the soft factor $\bZ$ of the usual 
(non-celestial) amplitudes (within the dipole approximation).\footnote{See, for instance, equation (2.9) in \cite{DelDuca:2011ae}} 
Therefore, $\bZ$ must reduce to the soft factor in \eqref{Msoft} for the most-subleading-color case, which is indeed the case.



\section{Possible corrections due to non-dipole terms}
\label{sec:nondipole}

In Section~4 we we worked under the dipole approximation assumption, that is, by only considering the terms 
in the first line of \eqref{fullGamma} to the soft anomalous dimension matrix $\bGam$. However, we know that 
there are corrections to it starting at 3-loops, as originally hinted in \cite{Caron-Huot:2013fea} and then explicitly 
shown in \cite{Almelid:2015jia}. In this section we want to consider the effect of those corrections.

Consider the 3-loop corrections computed in \cite{Almelid:2015jia}, 
\bea
\Delta_n^{(3)}(\{\rho_{ijkl})&=&16 f_{abe}f_{cde}\left\{\sum_{1\leq i<j<k<l\leq n}\left[\bT_i^a\bT_j^b\bT_k^c \bT_l^c{\cal F}(\rho_{ikjl},\rho_{iljk}
\right.\right.\cr
&&\left.\left.+\bT_i^a\bT_k^b\bT_j^c\bT_l^d{\cal F}(\rho_{ijkl},\rho_{ilkj})
+\bT_i^a\bT_l^b\bT_j^c\bT_k^d{\cal F}(\rho_{ijlk},\rho_{iklj})\right]\right.\cr
&&\left.-(\zeta_5+2\zeta_2\zeta_3)\sum_{i=1}^n\sum_{1\leq j<k\leq n,j,k\neq i}\{\bT_i^a,\bT_i^d\}\bT_j^b\bT_k^c\right\}\;,
\eea
where $\rho_{ijkl}=(s_{ij}s_{kl})/(s_{ik}s_{jl})$ are the conformally invariant cross-ratios,
\bea
{\cal F}(\rho_{ijkl},\rho_{ilkj})&=&F(1-z_{ijkl})-F(z_{ijkl})\cr
z_{ijkl}\bar z_{ijkl}&=&\rho_{ijkl}\;,\;\;\;
(1-z_{ijkl})(1-\bar z_{ijkl})=\rho_{ilkj}\cr
F(z)&=& {\cal L}_{10101}(z)+2\zeta_2[{\cal L}_{001}(z)+{\cal L}_{100}(z)]\;,
\eea
and the ${\cal L}_{...}(z)$ functions are Brown's single-valued harmonic polylogarithms. 

In the case of the 4-point function, we obtain 
\bea
\Delta_4^{(3)}(s/t)&=& 16 f_{abe}f_{cde}\left\{\bT_1^a\bT_2^b\bT_3^c\bT_4^d {\cal F}\left(\frac{u^2}{s^2},\frac{t^2}{s^2}\right)
+\bT_1^a\bT_3^n\bT_2^c\bT_4^d{\cal F}\left(\frac{s^2}{u^2},\frac{t^2}{u^2}\right)\right.\cr
&&\left.+\bT_1^a\bT_4^b\bT_2^c\bT_3^d{\cal F}\left(\frac{s^2}{t^2},
\frac{u^2}{t^2}\right)-(\zeta_5+2\zeta_2\zeta_3)\sum_{i=1}^n\sum_{1\leq j<k\leq n,j,k\neq i}\{\bT_i^a,\bT_i^d\}\bT_j^b\bT_k^c\right\}\cr
&=&16 T_1^aT_2^bT_3^cT_4^d\left\{f_{abe}f_{cde}{\cal F}\left(\frac{u^2}{s^2},\frac{t^2}{s^2}\right)
+f_{ace}f_{bde}{\cal F}\left(\frac{s^2}{u^2},\frac{t^2}{u^2}\right)\right.\cr
&&\left.+f_{ade}f_{bce}{\cal F}\left(\frac{s^2}{t^2},\frac{u^2}{t^2}\right)\right\}\cr
&&-(\zeta_5+2\zeta_2\zeta_3)f_{abe}f_{cde}\left[\{\bT_1^a,\bT_1^d\}(\bT_2^b\bT_3^c+\bT_2^b\bT_4^c+\bT_3^b\bT_4^c)\right.\cr
&&\left.
+\{\bT_2^a,\bT_2^d\}(\bT_1^b\bT_3^c+\bT_1^b\bT_4^c+\bT_3^b\bT_4^c)\right.\cr
&&\left.+\{\bT_3^a,\bT_3^d\}(\bT_1^b\bT_2^c+\bT_1^b\bT_4^c+\bT_2^b\bT_4^c)
+\{\bT_4^a,\bT_4^d\}(\bT_1^b\bT_2^c+\bT_1^b\bT_3^c+\bT_2^b\bT_3^c)\right].\cr
&&\label{3-loopnondip}
\eea

We then note that there is a matching of the first term ({\em i.e.}, the nonconstant term in curly brackets) 
to the more general formula considered in \cite{Naculich:2013xa}, which had 3 independent functions 
$P_s,P_t,P_u$, if we make the identifications
\be
\frac{1}{N^3}P_s=-16 {\cal F}\left(\frac{u^2}{s^2},\frac{t^2}{s^2}\right)\;,\;\;\;
\frac{1}{N^3}P_u=+16{\cal F}\left(\frac{s^2}{u^2},\frac{t^2}{u^2}\right)\;,\;\;\;
\frac{1}{N^3}P_t=-16 {\cal F}\left(\frac{s^2}{t^2},\frac{u^2}{t^2}\right).
\ee

Being so, we can import the calculation from there, and find 
\be
\Delta \bGam^\Three =
{1 \over N^3} \left( \begin{array}{cc} a & b \\
c & d
\end{array}
\right)
\ee
where 
\bea
a &=& 
\left(
\begin{array}{ccc}
 0 & 2 N (3 {P_t}-{P_s}-2 {P_u}) & 2 N (2 {P_u}- 3 {P_s}+{P_t}) \\
 2 N ({P_s}+2 {P_t}-3 {P_u}) & 0 & 2 N (3 {P_s}-2 {P_t}-{P_u}) \\
 2 N (3 {P_u}-2 {P_s}-{P_t}) & 2 N (2 {P_s}-3 {P_t}+{P_u}) & 0 \\
\end{array}
\right)\,,
\cr
b &=&
\left(
\begin{array}{ccc}
 8 ({P_t}-{P_s}) &
  2 N^2 ({P_t}-{P_s} ) + 4( P_u- {P_s}) 
& 2 N^2 ( P_t - P_s ) + 4 (P_t - P_u) \\
 2 N^2 ({P_s} -{P_u} )  + 4 ({P_s}- {P_t}) 
& 8 ({P_s}-{P_u}) 
& 2 N^2 ({P_s} -{P_u}) + 4 (  {P_t}- {P_u}) \\
 2N^2  (P_u-{P_t} ) + 4 ( {P_s}- {P_t}) 
& 2N^2 ({P_u}-{P_t} ) +4 (P_u - {P_s}) 
& 8 ({P_u}-{P_t}) \\
\end{array}
\right)\,,
\cr
c &=&
\left(
\begin{array}{ccc}
 2 ({P_t}-{P_s}) & (N^2+2) ({P_s}-{P_u}) & (N^2+2) ({P_u}-{P_t}) \\
 (N^2+2) ({P_t}-{P_s}) & 2 ({P_s}-{P_u}) & (N^2+2) ({P_u}-{P_t}) \\
 (N^2+2) ({P_t}-{P_s}) & (N^2+2) ({P_s}-{P_u}) & 2 ({P_u}-{P_t}) \\
\end{array}
\right)\,,
\cr
d &=&
\left(
\begin{array}{ccc}
 6 N ({P_s}-{P_t}) & 0 & 0 \\
 0 & 6 N ({P_u}-{P_s}) & 0 \\
 0 & 0 & 6 N ({P_t}-{P_u}) \\
\end{array}
\right)\,.
\eea

An interesting observation is that, for the most-subleading color part of this first term in (\ref{3-loopnondip}), $a=d=0$, so 
\be
\Delta \bGam^\Three_{\rm most-sub} =
{1 \over N^3} \left( \begin{array}{cc} 0 & b \\
c & 0
\end{array}
\right)\;,
\ee
which is the same block form that we have for $\bGam^\One_{\rm most-sub}$. However, the form of the
matrices $b$ and $c$ are different from the ones in $\bGam^\One_{\rm most-sub}$, and the two matrices 
do not commute (they just act in the same 
way, switching between single traces and double traces in the trace basis). This implies, for instance, that 
these 3-loop corrections to the KZ equation are rather non-trivial since they would involve new extra terms 
in the commutator expansion in \eqref{BCHexpansion}.

For our purposes, when computing the contributions of these corrections to celestial amplitudes, the 
important observation is that $\Delta_4^{(3)}$ is only a function of $r=-s/t$ (and more generally, 
$\Delta_n^{(3)}$ is a function conformal invariant cross ratios only), so its contribution to the 
celestial amplitude is trivially found since it just contributes to the ${\cal F}_L(r,\epsilon)$, that is, 
it becomes a simple constant that comes out of the integral over $w=-t$ in \eqref{eq:f}.

For completeness, note that considering only the most-subleading contribution of the  term 
$(\as^3/\eps) \Delta \bGam^\Three \ket{A^\Zero} $ ({\em i.e.}, the non-constant part in \eqref{3-loopnondip}),
when acting on the tree level amplitude
\be
\ket{A^\Zero}= - {4iK \over s t u} 
\left(
\begin{array}{c}
u \\ t \\ s
\end{array}
\right) \;,
\label{treelevel}
\ee
one obtains 
\be
\left(
\begin{array}{c}
\Delta A_4^\Threethree \\
\Delta A_5^\Threethree \\
\Delta A_6^\Threethree \\
\end{array}
\right)
=
-{8i  K\over \eps}  
{ [ (u-s) P_t + (s-t) P_u + (t-u) P_s  ]  \over 
s t u }  \ones 
+ \cO\left( \eps^0 \right) \,.
\ee

Note also that, as explained in \cite{Almelid:2015jia}, in the Regge limit $r=-s/t\rightarrow \infty$, $\Delta_4^{(3)}$ 
becomes a constant matrix (although still not one that commutes with the dipole terms).

\section{Conclusions}

In this paper we have considered the IR divergences of celestial amplitudes in gravity and in gauge theories, 
in particular the most-subleading-color part of the latter. We have found that the simplifications that occur in the IR structure of the 
most-subleading gluon amplitudes, responsible for their close connection with gravity, also translate into a 
simple form of the celestial correlators yielding a simple exponentiation and factorization.  The fact that the 
soft factor $\bZ$ in the most-subleading-color case is a function of conformal cross-ratios only plays a 
crucial role in the simple form of the celestial soft factorization.

Gravity amplitudes were found to have an exponential IR factor acting on the hard celestial part. 
If the dipole conjecture is assumed to be true, (such as in the case of ${\cal N}=4$ SYM where it {\em is} true), 
we were able to compute exactly the IR divergent factor acting on the hard part for the most-subleading celestial gluon amplitudes, as well 
as to reduce the two most divergent in $\epsilon$ terms to the calculation of the one-loop amplitude $A^{(1,1)}$, 
just like in the gravity case. 

Also, under the dipole approximation, for the all-order in $N$ soft factor $Z$ in terms of the cusp anomalous 
dimension $\gamma(\a)$, we have found a generalization of the 
Knizhnik-Zamolodchikov (KZ) equation, reducing to the KZ equation in the case of the most-subleading celestial gluon amplitude
in the Regge limit, as well as for the full amplitude at large $N$. Finally, we have also considered the 3-loop 
corrections due to non-dipole terms, and how they impact the IR divergent terms.

To conclude, we would like to close with a few observations and further directions:

\begin{itemize}
\item The KZ equation in the WZW model is a consistency relation as a consequence of the presence of a null 
state in the theory. Since the KZ equation is not satisfied here in general, suggesting the absence of null states, 
but {\em it does} arise in the large cross-ratio limit $r\to \infty$, this seems to imply that a null state does emerge 
in this particular limit. Also, as mentioned at the end of section 4, there could be other kinematical regions where 
the KZ equation is satisfied and it would definitely be interesting to explore this further.
 
\item Possible further directions include: (a) compute the subleading contributions to the Regge limit; 
(b) include $\Delta \Gamma_n$ corrections to the dipole term appearing at four loops; 
(c) consider the Regge limit of the full gluon amplitude (not just most-subleading part), using the 
results of \cite{DelDuca:2011ae};
(d) search for a non-trivial simplification also in the large $N$ limit; 
(e) consider the effect of the all-loop conjectured 
connection between ${\cal N}=4$ SYM and ${\cal N}=8$ supergravity in the Regge limit in the recent paper 
by S. Naculich \cite{Naculich:2020clm}.

\end{itemize}

\section*{Acknowledgements}


We especially thank Hern\'an Gonz\'alez for many 
discussions and collaboration in the initial stages of this project. 
We also would like to thank Steve Naculich, and Howard Schnitzer for discussions and observations on the final manuscript. 
 The work of H.N. is supported in part by CNPq grant 301491/2019-4 and FAPESP grants 2019/21281-4 and 2019/13231-7. 
H.N. would also like to thank the ICTP-SAIFR for their support through FAPESP grant 2016/01343-7. The work of F.R. has 
been supported by FONDECYT grants 11171148 and 1211545 as well as by ANID PIA Anillo ACT/210100. 
C.R. is supported by Proyecto Ingenier\'ia 2030 de CORFO, 
grant 14ENI2-26865.

\bibliography{celestial-refs-mostsubleading}
\bibliographystyle{utphys}

\end{document}